\documentclass[letterpaper]{article} 
\usepackage{aaai2026}  
\usepackage{times}  
\usepackage{helvet}  
\usepackage{courier}  
\usepackage[hyphens]{url}  
\usepackage{graphicx} 
\usepackage{natbib}  
\usepackage{caption} 
\usepackage{algorithm}
\usepackage{algorithmic}

\usepackage{dblfloatfix}

\usepackage{multirow}
\usepackage{tabularx}
\usepackage{booktabs}
\usepackage{amsmath}

\usepackage{newfloat}
\usepackage{listings}
\DeclareCaptionStyle{ruled}{labelfont=normalfont,labelsep=colon,strut=off} 
\floatstyle{ruled}
\newfloat{listing}{tb}{lst}{}
\floatname{listing}{Listing}
\title{Enabling the 1:1 Promise: \\
AI Enabled Fairness for Indonesia's Retail IPO Allocation
}

\author {
    Anchal Gupta\textsuperscript{\rm 1}, Gleb Papyshev\textsuperscript{\rm 2}, James T Kwok\textsuperscript{\rm 1}
}
\affiliations {
    \textsuperscript{\rm 1}The Hong Kong University of Science and Technology\\
    \textsuperscript{\rm 2} Lingnan University \\
    anchal@ust.hk, glebpapyshev@ln.edu.hk, jamesk@cse.ust.hk
}

\usepackage{bibentry}

\begin{document}

\maketitle

\begin{abstract}
Indonesia's OJK Circular, effective November 2025, requires that half of every IPO's pooled allocation go to retail investors. Uniquely amongst major capital markets, the regulation places all non-institutional participants, regardless of wealth, into a single competing pool rather than separating them by monetary thresholds. The rule is structurally sound; enforcement and measuring effectiveness require measures beyond rule-based detection alone. Well-resourced participants operating within the retail pool can distribute bids across networks of synthetic retail accounts, each individually compliant, to capture the retail quota undetected by standard concentration heuristics. Without a mechanism to detect such networks before allotment closes, the statutory guarantee of equal access becomes procedurally hollow: the quota is formally reserved for retail households whilst the economic benefit flows elsewhere. We present a framework identifying such nominee networks using eleven statutory fields already collected by Indonesia's e-IPO infrastructure, requiring no new surveillance powers or personal data. The framework maps account connections to identify hidden networks (Louvain), learns behavioural patterns signalling coordination by mining association rules (Apriori), and scores each network's likelihood of being a coordinated evasion ring (Random Forest). Evaluated on synthetic data bids ($n=154,000$) across 25 IPO offerings, calibrated against published e-IPO statistics and public disclosures, the framework serves two functions. Pre-allotment exclusion of nominee networks redistributes shares to genuine retail households, reducing allocation inequality by 63 to 66\% per offering (Gini). As supervisory intelligence, the model correctly separates nominee networks from genuine retail households in approximately 98 out of 100 cases (weighted AUROC $\approx 0.98$) even as actors shift tactics across offerings. False positives for kinship clusters are suppressed by design. This paper demonstrates that when a Global South regulator prioritises genuine inclusion over administrative convenience, algorithmic enforcement is not merely useful but demonstrably effective.
\end{abstract}


\section{Introduction}
What do you do when a country of 277 million
people~\cite{united_nations_population_fund_indonesia_2025}, with fast-growing
retail investor populations~\cite{gunawan_communication_2023}, decides its
capital markets must serve everyone equally not just those with institutional
access~\cite{cull_microfinance_2009, beck_access_2008}? In November 2025,
Indonesia answered this question. Otoritas Jasa Keuangan (OJK) Circular Letter
No. 25/SEOJK.04/2025~\cite{otoritas_jasa_keuangan_seojk_2025, fajarwati_new_2025,
abnr-counsellors_at_law_ojk_2026} restructured the Initial Public Offering (IPO)
allocation from a 1:2 retail-to-institutional ratio into a mandatory 1:1 parity,
collapsing three categories of investor (institutional, mass affluent, and
retail) into two: institutions and everyone else. This democratic market
experiment creates both an opportunity and a paradox. The opportunity is a
statutory mechanism for equitable wealth participation~\cite{azimli_impact_2023,
huang_initial_2022}. The paradox is that the mandate requires enforcement at a
scale and complexity that rule-based checks alone are not designed to address. Aforementioned checks evaluate each bid submission in isolation against the statutory 10\% per-account ceiling, confirming individual compliance; they cannot, by design, detect coordination distributed across many accounts that each remain within threshold when examined separately.

Prior to this reform, the Indonesian primary market suffered from severe
structural asymmetries. Institutional actors routinely engaged in coordinated,
concentrated bidding, often utilising fragmented nominee accounts to artificially
capture IPO allocations and manipulate post-listing price discovery. This
structural asymmetry is widely recognised in Indonesian domestic policy discourse
as \textit{'goreng saham'}(market manipulation)~\cite{gibran_saham_2025},
prompting need to have a sustained intervention from regulatory bodies to protect
retail participants. The structural asymmetry of the pre-reform regime reached
acute visibility in April 2025, seven months before SEOJK 25/2025 took effect, when Fore Coffee's IDX listing attracted 114,873 retail applicants and was oversubscribed 200 times under the then-governing 1:2 retail-to-institutional ratio, rendering allocation a zero-sum contest in which institutional capture was maximally incentivised~\cite{enerva_indonesian_2025, reuters_indonesian_2025, fore_coffee_east_2025}. The introduction of the 1:1 parity mandate in November 2025 was not a spontaneous policy shift but the culmination of a sustained macroeconomic leadership directive to restore retail confidence, with Fore Coffee's listing standing as the most visible illustration of the inequality the reform was designed to address.

The 2025 reform resolves the quota problem whilst generating a distinct enforcement one. Under OJK Circular 25/2025, no single investor's cumulative order may exceed 10\% of the total centralised allotment value, a per-account ceiling rather than a per-category one. A coordinated network of ten nominee accounts, each bidding to this individual maximum, can therefore absorb the entire 50\% retail quota whilst every account remains technically compliant. To OJK's standard ledger, these accounts appear as independent retail households; only their relational geometry reveals coordination. This regulatory architecture, whilst structurally sound in intent, engenders an enforcement dichotomy, by reserving 50\% of each offering for retail participants, the mandate creates material incentives for institutional actors to distribute large bids across networks of synthetic retail identities (SIDs), capturing the retail allocation whilst evading the per-account threshold entirely

OJK's regulation seeks to establish an equitable allocation environment by
conferring half of the pooled IPO shares upon all non-institutional participants
and capping any single investor, whether High Networth Individual (HNI) or otherwise, at 10\% of the total offer. This represents a departure from decades of structural asymmetry in which HNIs could concentrate allocations without constraint. Effective enforcement of this regime, however, requires surveillance capabilities extending beyond rule-based threshold checks; specifically, a system capable of identifying coordinated nominee networks across thousands of simultaneous IPO bids. If that surveillance is poorly calibrated, it will misclassify legitimate households as institutional proxies, penalising the very participants the regulation was designed to protect. A system that generates systematically higher false positive rates for the demographic groups a regulation was designed to protect constitutes, in this sense, an inequitable enforcement instrument; and a fairness regulation operationalised through such an instrument is self-defeating by design, reproducing through automated screening the distributional harms the statutory mandate was constructed to remedy. 

This paper demonstrates that Indonesia's allocation framework, structurally sound as a matter of policy design, can be made measurably functional through fairness-constrained AI enforcement, offering a transferable governance architecture from an emerging market regulator where primary market access is a first-order development priority and a principal channel for household wealth formation.

This paper contributes to the literature on algorithmic fairness and primary market
regulation across four dimensions. 
First, it furnishes a measured comparison, through a controlled simulation dataset, between standard rule-based threshold checks and the relational detection approach adopted here; the former identifies fewer than 5\% of coordinated nominee-layering networks, whilst the latter achieves near-complete separation, converting a previously theoretical surveillance gap into a documented empirical one.

Second, it adapts a privacy-preserving graph cluster detection framework, combining community partitioning, behavioural rule extraction and ensemble classification, applied exclusively to eleven statutory fields~\cite{otoritas_jasa_keuangan_seojk_2025}, all of which are collectable under standard centralised IPO platform architecture. No Indonesia-specific infrastructure is prerequisite for replication; the schema operates within existing regulatory data obligations without recourse to expanded surveillance powers.

Third, it utilises a fairness constraint at the graph partitioning stage, one
that protects demographically coherent household clusters from systematic
misclassification. The procedural mechanics are detailed in Section~\ref{sec:db}.

Fourth, it frames the enforcement paradox examined here as an instance of a structural condition rather than a jurisdiction-specific irregularity. Any capital market combining a statutory retail allocation quota, a unified competitive pool and a per-account rather than per-category concentration ceiling confronts the same inferential problem: individual compliance carries no information about collective coordination. The framework therefore offers a transferable enforcement approach for regulators in which allocation mandates exceed the protective capacity of account-level verification.

The remainder of this paper is organised as follows. Section 2 examines the role of primary capital markets as instruments of household wealth formation in emerging economies. Section 3 draws on film distribution and agricultural supply chains to illustrate how cumulative, seemingly federated but individually compliant actions can be an indicator of systemic market distortion and what it reveals about the need for equitable access enforcement. Section 4 describes the proposed detection framework, covering the statutory data schema, and two use cases: retail investor protection and regulatory strategic intelligence, which demonstrates how this framework can be useful at both individual and macro level. Section 5 addresses limitations and offers concluding remarks.

\section{Primary Markets, Retail Access, and the Enforcement Problem}

Primary equity markets (also referred as IPO) in emerging economies occupy a structurally distinct position from their counterparts in advanced financial systems: they function simultaneously as instruments of corporate finance, advancing the government's political agenda of privatisation~\cite{jones_share_1999} and as one of the few institutionally accessible channels through which household capital can participate in formal wealth creation~\cite{booth_ownership_1996}. Understanding why Indonesia's enforcement problem is both specific and consequential requires situating it within the broader context of what primary market access means for retail households in these economies, and why allocation mechanism design, rather than quota size alone, determines whether that access is substantive or illusory.

\subsection{IPOs as Wealth Formation Instruments in Emerging Economies}
Governments across emerging markets have recognised that public offerings represent a primary point of entry into equity ownership for non-institutional investors~\cite{menon_ravi_2014, hudaefi_exploring_2023}. Increased listing activity expands the tax base, liquidity in secondary markets and strengthens corporate governance norms. This generates employment and income effects that feed into broader distributional outcomes~\cite{zougarh_impact_2024, nnakee_examining_2024, jones_share_1999, booth_ownership_1996}. The World Bank's IPO-Plus framework explicitly frames public equity as a mechanism for promoting widespread, rather than concentrated, participation in privatisation and wealth creation~\cite{goldberg_ipo-plus_1999}.

This research focuses on IPOs rather than secondary markets because the IPO's single clearing event is the only point in the equity lifecycle where allocation parity is amenable~\cite{Qian_Ritter_Shao_2024}; secondary market trading is continuous, price-driven, and subject to execution asymmetries that no allocation rule can resolve, whereas the centralised e-IPO platform admits a defined enforcement boundary within which the 1:1 mandate can be operationalised.

These distributional benefits are, however, conditional on how shares are allocated, not merely on how many shares are reserved for retail investors~\cite{huang_initial_2022}. In practice, without explicit protective rules, large financial institutions capture the majority of IPO shares because they have better information~\cite{lin_ipo_2007}. For ordinary households, an IPO is often one of the only accessible opportunities to invest in a company before it begins trading publicly, making fair access a question of economic inclusion rather than financial regulation alone~\cite{menon_ravi_2014}. Evidence from India shows that when allocation rules are well designed, retail investors do receive better access to the most attractively priced offerings~\cite{wadhwa_reallocation_2018}. Indonesia's 2025 reform takes this principle further than many jurisdictions have attempted.

\subsection{Indonesia's Retail Parity Mandate}\label{sec:regulation}

OJK Circular 25/2025 mandates 1:1 parity within a unified competitive pool
without monetary separation, meaning high-net-worth individuals and institutions
compete directly against retail households under identical 10\% per-investor
caps, with no hard monetary floor excluding them from the retail
tranche~\cite{abnr-counsellors_at_law_ojk_2026,
otoritas_jasa_keuangan_seojk_2025, prima_wirayani_indonesia_2025}. We selected four jurisdictions on two criteria: India as the most comparable emerging economy by population scale and retail investor growth trajectory, and Hong Kong, the United States and the United Kingdom as three of the most frequently cited benchmark jurisdictions in comparative capital markets regulatory scholarship, with Hong Kong and the United States additionally ranking amongst the highest-volume IPO markets globally in 2025. A comparative review of these four major markets, specifically India~\cite{bbc_ipo_2025, groww_ipo_2025, banerjee_oxford_2016, sebi_india_sebi_2025, karim_anani_2025_2025}, Hong Kong~\cite{enoch_yiu_hong_2025}, the United States~\cite{finra_5131_2025} and the United Kingdom~\cite{securities_industry_and_financial_markets_association_2026_2026}, indicates that no comparable jurisdiction combines statutory retail parity, structural pool integration and the absence of monetary floors separating wealthy participants into a distinct tranche. In each case, wealthy individuals and institutions retain access to dedicated allocation channels, whether through hard monetary thresholds, guaranteed institutional quotas or the absence of any formal retail reservation requirement.On other hand, Indonesia allows them to compete in the same centralised e-IPO pool under these common rules.

This design is unusual when viewed against other large markets as Indonesia combines all three design elements simultaneously: a statutory 50\% retail quota, a unified competitive pool in which high-net-worth individuals and institutions face identical per-account caps alongside retail households, and the absence of hard monetary floors that would segregate wealthy participants into a separate channel. Indonesia's configuration is structurally singular, and that singularity in policy design needs a customised solution to ensure intent meets implementation.

This ambitious design creates a novel enforcement challenge: a rule-based system that checks each account individually cannot detect coordinated bidding spread across many accounts, which is generally done by institutional investors who have access to roadshows and better information~\cite{keloharju_distribution_2002}. Global South regulators like OJK, which face exactly these problems, remain under-represented in AI governance scholarship~\cite{roche_ethics_2023, correa_worldwide_2023}. The 1:1 parity rule creates a detection problem distinct from standard market surveillance. Because HNIs and institutions compete in the same pool as retail households, coordinated bidding across multiple accounts appears legitimate at the individual level: each bid respects the 10\% cap, each SID is validly registered. Only the pattern across accounts reveals coordination. This requires moving from transaction-level monitoring to network-level detection, a capability most existing surveillance infrastructure lacks.

\section{Cross-Disciplinary Patterns of Cumulative Market Distortion}

This section explores how distributed, individually compliant actions across different industries can aggregate into systemic barriers that displace small participants. By detecting these patterns in film distribution, agricultural supply chains and financial listings, we identify an analogous mechanism where assumed legitimate behaviour collectively subverts market diversity and equitable access.

In the motion picture industry, systemic exclusion often results from a phenomenon known as ``screen-stuffing''~\cite{hsu_jacks_2006}, where multiple cinema chains independently allocate their most lucrative showtimes to a single high-budget ``tentpole'' film. Although each individual theatre's decision represents a logical, compliant pursuit of localised profit, the aggregate effect across a metropolitan or national market is a ``cumulative displacement''~\cite{easley_differential_2016, barberis_style_2003} of independent and specialty films~\cite{hsu_jacks_2006}. Research into the survival of motion pictures suggests that a film's run length and screen count are heavily influenced by the presence of a ``nearest competitor'' showing the same title, particularly when theatres are under common ownership~\cite{einav_not_2010}. This creates a homogenised theatrical landscape functionally equivalent to allocation capture: independent films are denied theatrical presence not because any single theatre rejects them, but because the aggregated booking decisions of chains operating independently concentrate access on blockbusters alone.

A similar logic of aggregate exclusion is observable in agricultural supply chains, where the ``shelf-space'' of national retail giants acts as a centralised allotment pool. Where slotting fees are contractually explicit, nominee coordination is deliberately covert each synthetic account appears legitimate in isolation, rendering transparency-based regulation ineffective and necessitating the relational detection methods proposed here~\cite{reardon_rise_2003}. While no single contract may violate anti-trust thresholds, their collective weight creates a ``cumulative displacement'' of small-hold farmers who lack the volume-based leverage to compete for the same ``pooled'' shelf space. This process prioritises administrative convenience, reducing the number of supplier relationships for the retailer at the direct cost of market diversity and the economic inclusion of small producers.

Furthermore, this ``coordinated synthesis'' of distribution whether of creative media or physical produce acts as a barrier to entry for the ``small investors'' and intention of wealth distribution of the respective economies. This mechanism mirrors the coordinated market behaviour seen in finance, where the harm is not a single act of theft, but a collective narrowing of the market. By detecting how numerous small, compliant actions such as individual theatre managers or supermarket procurement officers prioritising a guaranteed, high-volume ``hit'' coalesce into a systemic distortion, we can better understand the displacement of diverse ``assets'' in any competitive ecosystem.

This pattern of aggregate distortion can also be found in IPOs. While machine learning for studying IPO listing performance, price manipulation, and insider trading is a commonly studied subject~\cite{pavon_perez_bias_2022, hilal_financial_2022, dou_ai-powered_2023, mclean_does_2016, gong_lie_2021}, less attention has been paid to distributed behaviour across multiple accounts that appears legitimate at the individual level but produces systemic distortion when aggregated~\cite{chen_parallel_2023, desai_finding_2024}. This distinction is critical for the OJK's pooled allocation mechanisms; the regulatory challenge is not identifying overt theft but rather detecting the ``cumulative displacement'' caused by well-resourced participants using nominee networks. Just as ``screen-stuffing'' uses legitimate business choices to crowd out independent film, synthetic retail accounts use compliant, small-scale bids to subvert the regulatory intent of democratised access.

Each of the three domains and the pattern of distributed, individually compliant
actions aggregating into systemic distortion appears across domains-from
theatrical exhibition to agricultural supply chains to, more recently,
cryptocurrency markets where Telegram-coordinated groups distribute manipulation
across hundreds of wallets below detection thresholds~\cite{bolz_real-time_2025}.
In each case, the analytical challenge is identical: the harm is not in any
single transaction but in the geometric relationship between them. A
transaction-level ledger can confirm that each bid respects the 10\% cap; it
cannot reveal that ten bids originate from a single coordinating entity. The
detection problem is, fundamentally, a graph problem. Nominee coordination
manifests as edge density shared funding accounts, correlated bid timing,
recurrent participant codes that becomes statistically legible only when
transactions are represented as nodes and their relationships as edges. This is
precisely why graph partitioning, rather than threshold-based heuristics, is the
appropriate analytical response. We now turn to the e-IPO statutory schema and
the Louvain-based detection framework~\cite{que_scalable_2015} that executes this
principle using only the  fields Indonesia already collects.

\section{The IPO Manipulation Detection Database}
\label{sec:db}

\begin{table*}[t]
\centering
\label{tab:ipo-database}
\begin{tabularx}{\textwidth}{l l X X}
\toprule
\textbf{Category} & \textbf{Field} & \textbf{Description} & \textbf{Analytical Utility} \\
\midrule
\multirow{3}{*}{Identifiers} 
& SID\_Investor\_ID & Single Investor Identification & Primary unit of analysis at investor level \\
& IPO\_ID & Unique offering identifier & Enables temporal sequencing and cross-offering persistence measurement \\
& RDN\_Account\_Hash & Hashed customer fund account number & Detects shared funding relationships \\
\midrule
Temporal Footprint 
& Bid\_Timestamp & Time of order submission & Enables timing-based pattern analysis \\
\midrule
\multirow{4}{*}{Order Details} 
& Requested\_Lots & Number of lots requested & Supports analysis of bid sizing patterns \\
& Bid\_Price & Price submitted in the bid & Enables price-related behavioural analysis \\
& Bid\_Amendment\_Flag & Binary indicator of amendment & Supports amendment intensity and resubmission pattern analysis \\
& Tranche\_Type & Investor pool classification & Supports tranche-level and pool-composition analysis \\
\midrule
\multirow{2}{*}{Demographics} 
& SID\_Age\_Days & Age of the investor account & Supports account-tenure analysis \\
& Investor\_Age\_Yrs & Investor age in years & Supports demographic variation analysis \\
\midrule
Geography 
& Kota & City of investor registration & Enables geographic clustering and regional behavioural profiling \\
\bottomrule
\end{tabularx}
\caption{IPO Manipulation Detection Database: Eleven statutory fields extracted from Indonesia's centralised e-IPO infrastructure, organised by analytical domain. Fields are restricted to those mandated under existing OJK regulatory documentation; no device identifiers, biometric data or external enrichment is included.}

\vspace{-6pt}
\end{table*}

For this research, we constructed a privacy-preserving analytical database using
eleven fields drawn from the transaction records generated through Indonesia's centralised e-IPO system~\cite{id_clear_eipo}. These fields were selected because they are sufficient for detecting relational and behavioural patterns relevant to allocation analysis, while avoiding unnecessary use of personal or sensitive information.

The schema excludes direct identifiers; quasi-identifiers (age, city) are retained at coarse granularity (years, municipal level) and subject to anonymity thresholds for outlier combinations, balancing fairness-constraint utility against re-identification risk. The database as outlined in Table 1 is therefore designed as a restricted research subset rather than a replication of the full operational platform dataset. It contains only the fields needed for analysis and excludes broader administrative, identity, and account information collected for operational or compliance purposes. Nominee coordination leaves statistical traces and timing entropy detectable through graph analysis without requiring invasive behavioural biometrics.

The database schema operationalises detection through five categorical domains:
identifiers,temporal footprint, order details, demographics and geography. The
identifier fields establish relational anchors through investor, order, and
participant codes. Temporal field records submission timing and sequencing,
enabling entropy analysis of coordinated behaviour, and order details capture
sizing and pricing decisions. Demographic fields provide account tenure and
investor age, supporting variance analysis across household structures. 
What distinguishes this schema is its deliberate restriction to fields already collected under Indonesia's existing e-IPO infrastructure across household structures, removing the need for expanded surveillance powers or invasive biometric collection. This parsimony is architecturally intentional: nominee coordination manifests through relational and statistical traces detectable via graph analysis without recourse to external ground truth or personally identifiable data. The schema thus balances analytical sufficiency against privacy minimisation, demonstrating that effective enforcement of the 1:1 parity mandate requires neither expensive data collection nor compromised regulatory legitimacy.

Community detection algorithms are structurally biased towards high-degree nodes~\cite{de_vink_measuring_2025}, which risks systematically flagging densely connected household clusters as suspicious. To counteract this, the framework embeds a fairness constraint at the partitioning stage. Each cluster receives a suspiciousness score, defined as the negated anomaly output of an Isolation Forest~\cite{hilal_financial_2022} model fitted on the statutory fields, augmented by a community risk score encoding each cluster's relational density from the Louvain partitioning step. Clusters whose demographic composition indicates multi-generational household bidding, specifically an inter-quartile range of investor age exceeding fifteen years, receive a score override of zero irrespective of their anomaly signal, excluding them from enforcement consideration entirely. The fifteen-year threshold reflects Indonesia's investor age structure, in which the dominant retail cohort ($\le 30$ years, 58\% of registered investors (KSEI 2024)) and their parental generation (41–60 years) produce natural inter-generational spreads well above this bound in genuine household clusters~\cite{kreager_indonesia_2008,tang_analysis_2024} , while nominee rings exhibit age homogeneity by construction. This constraint directly addresses the differential false positive risk that detection across demographic groups introduces rates~\cite{hes_machine_2025}, the mechanism is designed so that the households the regulation was constructed to protect cannot be systematically misclassified by the very instrument enforcing it.

\section{Database Use Cases}
This section describes the construction of the synthetic dataset and presents two use cases through which the detection framework is evaluated: retail investor protection at the pre-allotment stage and adaptive governance intelligence across successive offering cycles.

\subsection{Database Construction} \label{sec:5.1}

This database construction began by studying the fields and kind of information collected by Indonesia e-IPO and India~\cite{id_clear_eipo, groww_ipo_2025}. Publicly available e-IPO transaction records do not include field-level schemas or sample records. To ensure realistic data structure, we reconstructed the eleven-field schema from three sources: (i) OJK Circular 25/2025 regulatory filings specifying mandatory data elements for IPO verification; (ii) ID Clear technical documentation describing the e-IPO platform's data collection interface; and (iii) comparative analysis with India's UPI-based IPO System to infer the fields for retail equity offerings in emerging market centralised platforms~\cite{groww_ipo_2025}. These sources were cross-referenced to identify common mandatory fields across jurisdictions, then mapped to Indonesia's specific regulatory requirements. Synthetic data generation used MostlyAI's~\cite{krchova_improving_2025} sequential and relational modelling capabilities to preserve intra-record dependencies.

\begin{figure*}[!b]
    \centering
    \includegraphics[width=\textwidth]{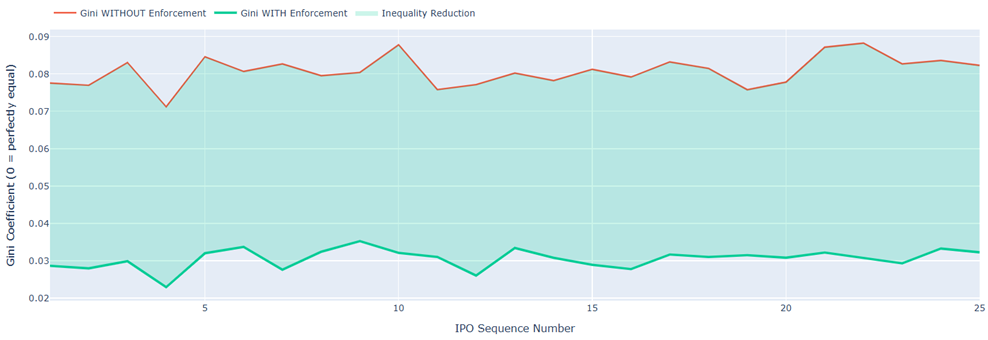}
    \caption{Allotment inequality across 25 synthetic IPO offerings, measured by the Gini coefficient. The upper series represents the distribution without enforcement; the lower series represents the distribution following nominee exclusion and lot redistribution.}
    \label{fig:placeholder}
\end{figure*}

The full dataset comprises 154,000 synthetic bids spanning 25 IPO offerings. The statistical parameters replicating in this dataset are drawn from three sources: marginal field distributions calibrated against published e-IPO statistics and the Fore Coffee 2025 offering disclosure, network degree distributions matched to confirmed goreng saham enforcement cases, and bid submission timing entropy aligned to Indonesian retail investor platform behaviour documented in prior empirical work. Use Case 1 operates on the full 154,000-bid dataset, where each unique investor-offering pair forms a node in the bidding graph. Of the 141,013 nodes, 24,961 share relational edges with at least one other account; Louvain community detection identifies coordinated clusters from within this connected subset.

The database supports two distinct regulatory functions. The first, detailed in Use Case 1, addresses pre-allotment enforcement: by screening clusters before each offering closes, the framework redirects misappropriated lots towards genuine retail households. The second, detailed in Use Case 2, addresses supervisory continuity: by tracking whether coordination signatures remain detectable as nominee behaviour evolves across offerings, the framework provides OJK with the institutional foresight to sustain enforcement over time. The two functions operate on related but distinct analytical subsets of the dataset, as described above, and are read in tandem rather than in isolation.

The synthetic data generation preserved empirically observed relational and temporal patterns through three validation layers. First, marginal distributions of each field (account age, bid timing, lot sizing) were calibrated against published e-IPO statistics~\cite{idx_closing_2025, ojk_capital_nodate} and the Fore Coffee 2025 offering disclosure~\cite{fore_coffee_east_2025}, ensuring demographic and behavioural realism. Second, network structure was validated against known \textit{goreng saham} enforcement cases~\cite{chevalier_manipulated_2025}, with synthetic nominee clusters replicating the degree distribution and clustering coefficients observed in confirmed coordination networks. Lastly, temporal patterns specifically bid submission entropy and Intraday clustering were matched to the timing signatures documented in Hudaefi et al.~\cite{hudaefi_exploring_2023} regarding Indonesian retail investor platform behaviour. The resulting dataset exhibits consistency with historical Indonesian IPO performance, providing face validity for the coordination detection task.

\subsection{Use Case 1: Retail Investor Protection}

This use case executes the 1:1 mandate at the transaction level: before each offering closes. Prior to each IPO offering's allotment finalisation, the framework generates a ranked exclusion list identifying nominee clusters, which OJK may then act upon to redirect mis-captured lots towards verified retail participants before settlement closes.

Nominee networks do not violate any individual rule. Each account is validly registered, each bid respects the 10\% concentration cap and each submission appears retail-compliant in isolation. The harm is structural, across representative offerings in this dataset, such networks account for approximately 15\% of all detected clusters whilst capturing 20\% of total applicant accounts.
As a baseline comparison, standard rule-based threshold checks applied to the same dataset successfully flagged fewer than 5\% of the confirmed nominee networks, highlighting the structural evasion problem.

Detection operates at the network level because the evasion is relational rather
than transactional. Louvain community detection partitions the bidding graph into clusters, and Isolation Forest then scores each cluster using nine features derived from the eleven statutory fields: cluster size, submission timing entropy, intraday burst density, fresh account ratio, investor age variance, cross-offering recurrence count, Louvain community risk score, temporal proximity amongst co-bidders, and lot-size coefficient of variation. The principal discriminating signals are bid sizing patterns, amendment behaviour and cross-offering recurrence count. Each cluster receives a suspiciousness score; those scoring above a calibrated threshold are designated nominee candidates and placed on the pre-allotment exclusion list. Kinship clusters are assigned a score of zero and excluded from the candidate set entirely, irrespective of their anomaly signal, in application of the fairness constraint described previously. This ensures that the detection mechanism does not penalise the participants it is designed to protect.

Figure~\ref{fig:placeholder} plots the Gini coefficient~\cite{lee_measuring_2025,
yesha_yadav_failure_2016} across all 25 offerings, with and without enforcement where a lower coefficient indicates more equal distribution of allocated shares amongst applicants and therefore greater alignment with the policy objective of the 1:1 mandate. The x-axis records the IPO sequence number, corresponding to the chronological ordering of IPO\_ID values in Table 1. The Gini coefficient without enforcement holds persistently between 0.075 and 0.088 across all 25 offerings. Under enforcement, it settles consistently between 0.025 and 0.033. The shaded band between them represents the inequality reduction delivered per offering: wide, stable, and uniform from the earliest to the latest IPO in the dataset. That reduction translates to over 1200 share lots per offering redirected towards genuine retail households, using only the statutory fields Indonesia's infrastructure already collects. The framework's output is a ranked exclusion list, a decision-support instrument, not an automated enforcement action. The determination to act upon it, and the timeline relative to settlement closure, remain institutional decisions requiring OJK process design.

The uniformity of the result across the full offering sequence indicates that the gain is not incidental; nominee displacement is a structural feature of the market, and pre-allotment screening, applied consistently, is a proportionate and durable corrective. 


\subsection{Use Case 2: OJK Adaptive Governance Intelligence}

OJK leadership, before committing to continuous enforcement across successive offering cycles, must resolve a question that pre-allotment screening alone cannot answer. This is not whether the framework detects coordination under current market conditions, but whether its discriminative capacity persists as coordinated actors modify their tactics incrementally over time. This use case addresses that question directly. It operates on a modified version of the 154,000-bid dataset described in Section 5.1, with feature engineering applied to simulate varying degrees of behavioural sophistication across the 25 offerings. The use case is designed to be read as a supervisory instrument rather than a one-time classification result.

The analytical pipeline operates in two stages. In the first stage, Apriori association rule mining~\cite{guidotti_designing_2021} is applied separately to three temporal cohorts: early offerings (IPOs one to eight), mid offerings (nine to sixteen) and late offerings (seventeen to twenty-five). Each IPO is treated as a transaction and each observable pattern in bidding behaviour, such as how many lots were requested, how often bids were amended and how frequently the same account appeared across multiple offerings, is treated as an item. The algorithm identifies which combinations of these patterns recur consistently amongst coordinated nominee clusters within each cohort.In the second stage, each identified pattern is converted into a yes-or-no feature: does a given cluster exhibit this pattern or not? The resulting feature set is passed to a Random Forest classifier~\cite{jo_learning_2023}, trained to distinguish coordinated nominee clusters from genuine retail households, using ground-truth labels assigned during synthetic data generation. The model is trained on the first eighteen offerings and tested on the remaining seven.

Validation is conducted under two protocols to confirm that performance does not depend on temporal proximity to the training window. The chronological protocol trains on IPOs one to eighteen and tests on nineteen to twenty-five, yielding an AUROC of 0.9793. The skip-gap protocol \cite{ieee_computational_intelligence_society_ieee-cis_2019,
michael_jahrer_safe_2018} trains on IPOs one to sixteen, deliberately omitting the two cohorts immediately preceding the test window. This ensures the model is evaluated on offerings temporally distant from its training data rather than merely adjacent, testing whether the framework has learnt durable coordination signatures or memorised surface regularities of the most recent offerings. The skip-gap AUROC is 0.9788. The 0.0005 deviation between protocols, both evaluated on an identical test set, is insufficient as a standalone independence test; robustness is more credibly evidenced by the widening-gap results and the Jaccard stability gradient reported above.

\begin{figure}[H]
\vspace{-4pt}
\centering
\includegraphics[width=\columnwidth,keepaspectratio]{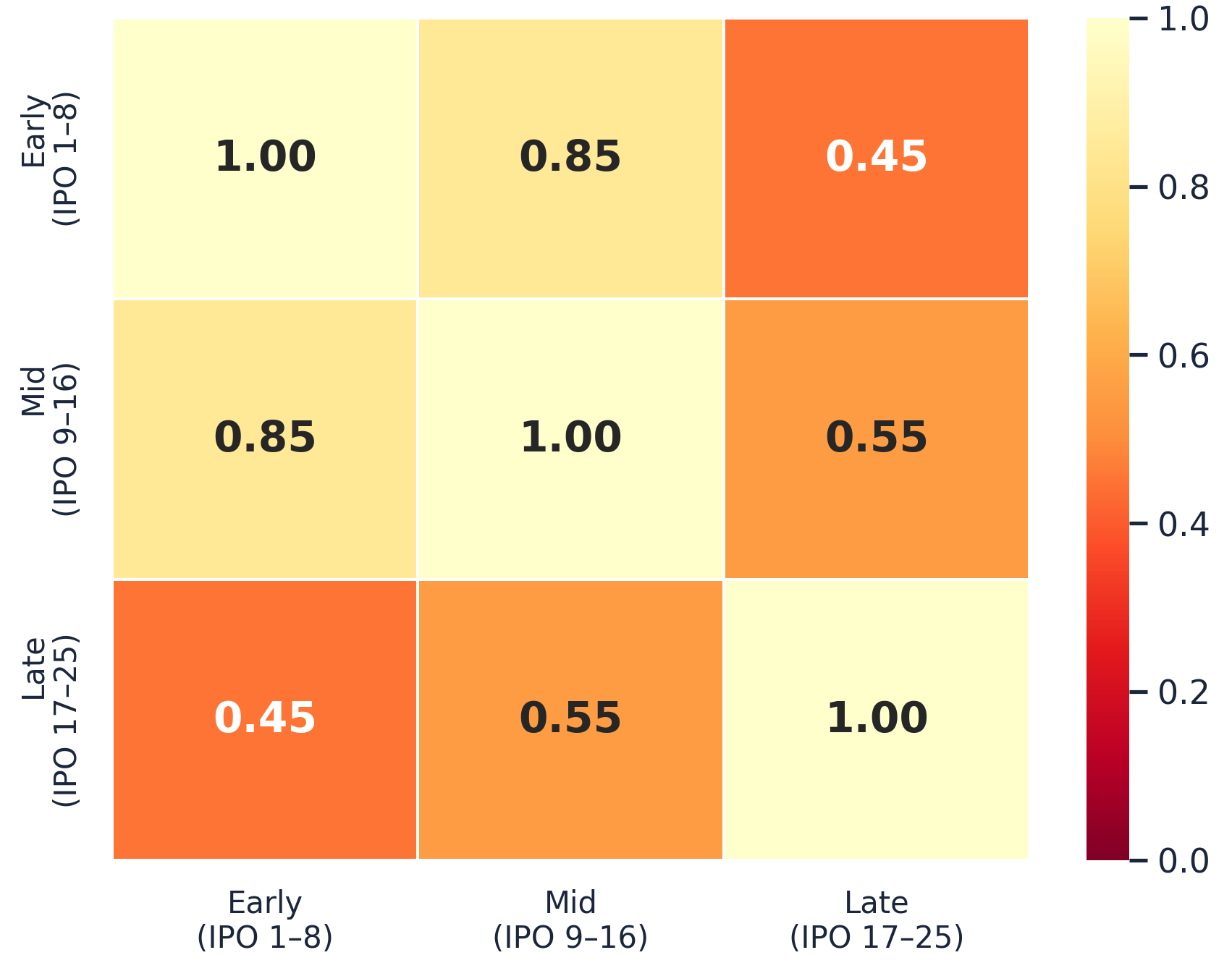}
\caption{Rule Migration Heatmap—Pairwise Jaccard similarity between dominant association rules extracted from early to late IPO cohorts. The gradient from 0.85 to 0.45 across sequence indicates incremental tactical adjustment rather than wholesale evasion, with the early-to-late score of 0.45 falling above the 0.40 adaptation threshold.}
\label{fig:rulemigration}
\vspace{-6pt}
\end{figure}

\begin{figure*}[!t]
    \centering
    \includegraphics[width=1\linewidth]{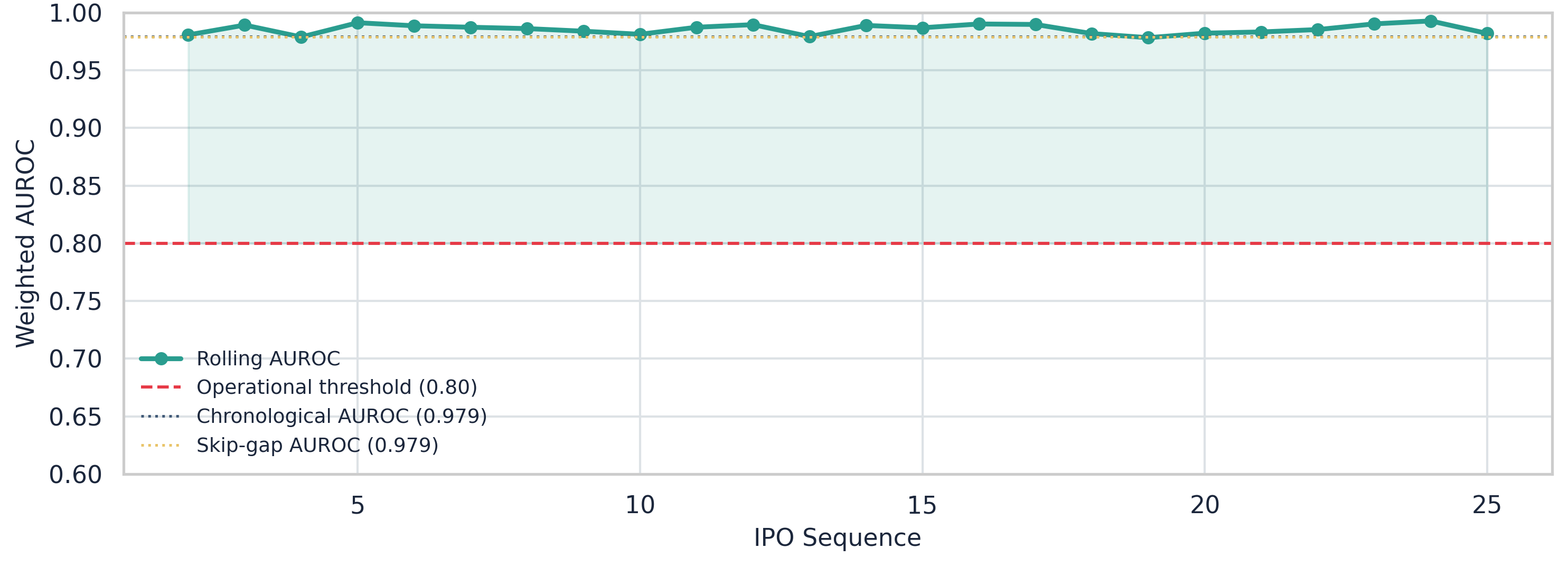}
    \caption{Use Case 2-Adaptive Governance Intelligence: Weighted AUROC across 22 rolling IPO windows. Dashed red line: 0.80 operational threshold. Dotted lines: chronological (0.979) and skip-gap (0.979) reference values. All windows remain above threshold; mean 0.9857, SD 0.0043.}
    \label{fig:rollingauroc}
\end{figure*}

The two figures produced by this pipeline answer distinct governance questions and are read in tandem. Figure~\ref{fig:rulemigration} addresses the prospective question: will performance be maintained as actors adapt? The rule migration heatmap plots Jaccard similarity~\cite{yang_llm_2025}, which measures the proportion of dominant coordination patterns that remain consistent between two cohort periods, across early, mid and late cohort pairs. A score of 1.00 indicates identical rule sets; a score of 0.00 indicates complete divergence. Jaccard similarity between early and mid cohorts registers 0.85, indicating substantial continuity in coordination tactics. Between early and late cohorts it falls to 0.45, indicating that a meaningful fraction of dominant signatures have been modified across the sequence whilst a structural core remains sufficiently stable to sustain detection. This gradient is the operationally significant signal: it indicates incremental tactical adjustment rather than wholesale evasion, and implies that periodic retraining at intervals of eight to ten offerings, using the three most recent cohorts, is a proportionate response to preserve discriminative capacity as surface patterns continue to shift.

Figure~\ref{fig:rollingauroc} addresses the retrospective question: is the model currently performing adequately? Across 22 rolling windows, each requiring a minimum of 200 training and 30 test observations, the weighted AUROC, computed as support-weighted one-versus-rest across four participant classes, averaged 0.9857 (SD = 0.0043), remaining consistently above the 0.80 operational threshold. A value approaching 0.98 indicates near-complete separation of genuine retail accounts from coordinated nominee clusters. Amongst the 5,609 genuine retail households examined in the test set, none was misclassified as a nominee ring, yielding a household false positive rate of nearly zero, against a target of below 5\%.

For a surveillance officer preparing a briefing for senior leadership, the two figures are read as follows. A stable AUROC band, remaining well above the 0.80 operational threshold, authorises commitment to the forthcoming offering cycle. A declining Jaccard gradient~\cite{yang_llm_2025}, irrespective of whether headline AUROC has yet deteriorated, is the forward indicator that a supervised retraining cycle is due. The capacity to act before performance erodes, rather than in response to deterioration already recorded, is what distinguishes a governance instrument from a detection tool. The framework therefore does not merely enforce the 1:1 mandate offering by offering; it provides OJK with the evidential basis to sustain that enforcement as market conditions evolve.

\subsection{Synthesis: From Enforcement to Equitable Access}
Taken together, the two use cases execute Indonesia's 1:1 parity mandate at both the transaction and the institutional level. The distributional gains recorded in Use Case 1 are not self-perpetuating. They are contingent on the supervisory architecture of Use Case 2 remaining calibrated as coordinated actors modify their tactics incrementally across offerings, as the declining Jaccard gradient makes plain. A regulator equipped with both capabilities can present to parliamentary oversight not merely that the framework detected coordination historically, but that it is structurally positioned to continue doing so as market conditions evolve. Indonesia's reform rests on the proposition that primary market access constitutes a channel for household wealth formation; what the two use cases establish, taken together, is that this proposition is technically realisable using only the statutory infrastructure Indonesia already possesses.

\section{Conclusion}
Justice, in the formulation offered by Indonesian moral philosophy, is realised through the structures that govern how opportunity is distributed amongst members of a society~\cite{fachri_keadilan_2019}; it is not possessed abstractly, but enacted relationally. The statutory obligation created by OJK Circular 25/2025 translates this principle into operational form. What this paper has demonstrated is that such translation is technically achievable without expansion of surveillance powers, without compromise of privacy minima, and without reproducing the distributional harms the regulation was designed to prevent.

Technique alone, however, is insufficient. Reflecting on IMF observation on financial surveillance~\cite{gabor_imfs_2015}, one cannot navigate by instruments alone when the terrain itself shifts beneath the vessel. The framework therefore incorporates temporal validation as integral architecture rather than afterthought, and the monitoring capability demonstrated in Use Case 2 transforms the database from an enforcement instrument into institutional memory. Regulators across markets at varying stages of allocation reform frequently proceed on theoretical intuition; this architecture offers a replicable evidence base for iterative policy design, enabling rule changes to be modelled against simulated behavioural response before crystallising into binding obligation.

The framework carries deliberate calibration choices that bound its scope. It is oriented towards institutional-scale nominee rings rather than extended family clusters, and the residual grey zone between kinship bidding and coordinated evasion warrants continued refinement as adversarial sophistication increases. Two implementation questions also fall outside this paper's scope: the procedural rights of excluded participants under Indonesian administrative law (Law No. 30 of 2014)\cite{indonesia_central_government_law_2014}, and the false positive rate under live deployment conditions where kinship and coordination signatures partially overlap. Both are directions for regulatory and empirical refinement.

The transferable insight of our framework can be summarised as where centralised, population-scale transaction infrastructure exists, effective algorithmic enforcement need not demand ever-expanding data collection. The consistent reduction in allotment inequality across all offerings, achieved using only eleven statutory fields already collected, suggests that data minimisation and enforcement efficacy are not traded off but jointly optimised through analytical method. This offers a replicable and transferable governance architecture for any market regulators, and a counterexample to the assumption that algorithmic fairness in financial surveillance requires compromise between detection performance and the protection of legitimate participants. The regulation promises equitable participation; the techniques of AI and machine learning demonstrate that such promises can be kept.

\section*{Acknowledgements}
This research was supported in part by the Research Grants Council of the Hong Kong Special Administrative Region (Grants 16202523 and HKU C7004-22G).

We also want to thank editorial team who have patiently guided us to get this paper camera-ready. 

\bibliography{IPO-BIB}

\end{document}